\documentclass[11pt]{article}
\usepackage[includeheadfoot,margin=1in]{geometry}
\usepackage{times}
\usepackage{cite}
\usepackage{amsmath,amsfonts,amssymb, booktabs}
\usepackage{mathtools}
\usepackage{enumerate}
\usepackage{verbatim}
\usepackage{commath}
    \usepackage[usenames]{color}
    \definecolor{plum}  {rgb}{.4,0,.4}
    \definecolor{BrickRed} {rgb}{0.6,0,0}
	\definecolor{DarkBlue} {rgb}{0,0,0.6}
\usepackage[plainpages=false,pdfpagelabels,colorlinks=true,linkcolor=BrickRed,citecolor=plum,urlcolor=DarkBlue]{hyperref}
\usepackage[mathcal]{euscript}
\usepackage{cleveref}

\begin{document}
\title{A Multi-Port Concurrent Communication Model for handling Compute Intensive Tasks on Distributed Satellite System Constellations}
\author{\Large Bharadwaj Veeravalli \\ \href{mailto:elebv@nus.edu.sg}{elebv@nus.edu.sg} \\ %
Department of Electrical and Computer Engineering,\\ National University of Singapore, 4 Engineering Drive 3, Singapore.}
\date{\today}
\maketitle
\begin{abstract}
We develop an integrated Multi-Port Concurrent Communication Divisible Load Theory (MPCC-DLT) framework for relay-centric distributed satellite systems (DSS), capturing concurrent data dissemination, parallel computation, and result return under heterogeneous onboard processing and inter-satellite link conditions. We propose a formulation that yields closed-form expressions for optimal load allocation and completion time that explicitly quantify the joint impact of computation speed, link bandwidth, and result-size overhead. We further derive deadline feasibility conditions that enable explicit sizing of cooperative satellite clusters to meet time-critical task requirements. Extensive simulation results demonstrate that highly distributable tasks achieve substantial latency reduction, while communication-heavy tasks exhibit diminishing returns due to result-transfer overheads. To bridge theory and practice, we extend the MPCC-DLT framework with a real-time admission control mechanism that handles stochastic task arrivals and deadline constraints, enabling blocking-aware operation. Our real-time simulations illustrate how task structure and system parameters jointly govern deadline satisfaction and operating regimes. Overall, this work provides the first analytically tractable MPCC-DLT model for distributed satellite systems and offers actionable insights for application-aware scheduling and system-level design of future satellite constellations.
\end{abstract}
%
\section{INTRODUCTION}
{A} significant change in the way sensing, communication, and computation are carried out in space has been brought about by the quick deployment of Distributed Satellite Systems (DSS) and large-scale low-Earth-orbit (LEO) constellations. Modern satellite networks increasingly rely on cooperative designs, where groups of satellites work together to perform compute-intensive activities, including processing Earth observations, managing broadband traffic, and gathering space-edge intelligence, rather than functioning as separate platforms. Selected spacecraft act as relay or coordination nodes in many of these systems, collecting data from nearby satellites via inter-satellite connections (ISLs/non-terrestrial networks) and processing it partially or completely before sending the findings to gateways or higher network levels \cite{DZhou2022,MGio2021}. Work in \cite{DZhou2022} explicitly discusses LEO satellites acting as relay nodes that collect and forward data to ground gateways in integrated satellite-terrestrial networks. 

Satellites may now maintain numerous simultaneous communication links with nearby nodes thanks to recent advancements in ISL technology, including as laser-based crosslinks, phased-array antennas, and digital beamforming. These capabilities are essential to developing multi-layer satellite network architectures and are currently utilized in planned and operating constellations like OneWeb, Starlink, and other non-terrestrial networks \cite{DZhou2022,MGio2021}. The requirements for effective task partitioning algorithms that simultaneously take into account heterogeneous processing capabilities, communication constraints, and result-aggregation overheads have also been brought to light by the increased interest in in-orbit and space-edge computing \cite{Mark2023}.

We will employ Divisible Load Theory (DLT) \cite{Bharadwaj1996Book}, which provides a mathematically rigorous framework for analyzing such task distribution problems when tasks can be arbitrarily partitioned. DLT has been widely used to derive optimal scheduling and load-allocation policies in distributed computing systems. Nevertheless, single-port or sequential communication models, which force artificial serialization on data transfers, are assumed by the majority of traditional DLT formulations. Relay satellites can use multi-port transceivers to transmit to and receive from several neighbors simultaneously, making the above-mentioned single-port assumptions increasingly impractical for modern DSS. Therefore, it is evident that DLT-based formulations under communication models reflecting the genuine parallelism available in modern satellite networks need to be reviewed.

In parallel, a recent work has proposed split learning frameworks tailored for LEO satellite networks to overcome the challenge of intermittent satellite-ground station connectivity\cite{ZLin2025}. However, the focus of this work is more on developing ML approach and bidirectional data exchange was out of scope. However, the applicability of this split learning model seems promising for a DLT-based approach. Another recent work on satellite-enabled collaborative intelligence—such as split learning\cite{WWu2023}, federated learning, and distributed inference—has emphasized the importance of bidirectional data exchange, where partial results or intermediate representations must be returned to a coordinating node.  Any realistic performance model must explicitly account for these result-return costs since they are not insignificant in practice. Furthermore, analytical models that can handle non-uniform processing and communication are required because of  the heterogeneity inherent in DSS and in edge computing platforms, which arises from different satellite payloads, orbital dynamics, and link circumstances\cite{Mark2023}. In a related context, the work reported in \cite{XZhang2024} attempts to jointly minimize delay and energy consumption under system resource and backlog constraints, developing an online distributed algorithm proven to achieve near-optimal performance in time-varying satellite edge computing networks.
Motivated by these observations, in this paper we develop a Multi-Port Concurrent Communication (MPCC)–enabled DLT formulation for DSS constellations. We focus on a canonical yet practically relevant topology in which each relay satellite forms a single-level star network with a set of neighboring satellites within ISL range. Under the MPCC model, the relay can distribute a divisible task to all neighbors concurrently and collect computed results in parallel. By applying the DLT optimality criterion, we derive closed-form expressions for optimal load fractions and makespan while explicitly accounting for heterogeneous processing speeds, heterogeneous ISL rates, and result-return overhead. The resulting formulation provides valuable insights into the fundamental performance limits of DSS and offers a tractable tool for the design and analysis of cooperative satellite (real-time) computing architectures \cite{Mark2023}.
\vspace{-0.4cm}
\subsection{Relevant Literature}
Now we shall present some of the very relevant DLT literature. For several years, DLT has been used as a fundamental framework to analyze the best way to divide the tasks in distributed computing systems with heterogeneous communication and computation. The theoretical foundations of divisible tasks, equal-finish-time optimality, and recursive load allocation for tree and star networks under sequential communication models were developed by early foundational publications in \cite{Bharadwaj1996Book, Robertazzi2003}. Some recent research has expanded DLT to cloud and edge computing contexts by including heterogeneous processors, energy limitations, and dynamic arrivals, building on these foundations \cite{Drozdowski2020,Li2021}. Other recent works targeting  specific applications include,  \cite{Joanna2023} and \cite{Jed2020}. In\cite{Joanna2023}, authors applied DLT paradigm for filtering and processing large data streams in intelligent data acquisition systems, whereas in \cite{Jed2020}, the authors clearly elicited a trade-off study between time and energy metrics in a hierarchical memory system in distributed platforms.  DLT has lately been reexamined in the context of satellite and space-edge systems, where onboard processing constraints and inter-satellite link unpredictability are considered as important factors when it comes to computation offloading for satellite edge computing \cite{QTang2021}. However, concurrent communication and result-return overheads have not been attempted so far on platforms employing multi-port communication abilities. Despite these advances, analytically tractable DLT models that explicitly capture multi-port concurrent communication for DSS remain largely unexplored, motivating the present work.
\vspace{-0.4cm}
\subsection{Objectives and scope of this work}
The objective of this work is to develop and analytically characterize a DLT framework for DSS constellations under an MPCC model that accurately reflects the capabilities of modern inter-satellite links. Focusing on a relay-centric single-level topology, the scope of the paper is to derive \textit{closed-form expressions for optimal load fractions and completion time while explicitly accounting for heterogeneous processing speeds, heterogeneous ISL rates, and non-negligible result-return overheads}. Beyond static task execution, the framework is extended with an admission control mechanism to support real-time task arrivals under deadline constraints, enabling the study of blocking behavior and operating regimes in dynamic DSS environments. The proposed formulation is intended to serve as both a performance benchmark and a design tool for cooperative satellite computing architectures. To complement the theoretical analysis, the derived results are systematically evaluated through rigorous simulation-based experiments, examining the impact of system heterogeneity, result-size ratios, constellation scale, and arrival intensity on latency, feasibility, and admission performance.\\
The organization of this paper is as follows. In Sections 2 and 3, we present the details MPCC-DLT formulation and analysis followed by rigorous performance evaluations in Section 4. In Section 5, we present real-time admission control for MPCC-DLT in DSS and finally, in Section 6, we conclude the work by highlighting important contributions and plausible extensions to this work. 
\vspace{-0.45cm}
\section{System Model - Bridge to MPCC–DLT Formulation}
We consider a DSS constellation segment in which a relay satellite forms a single-level star topology with $N$ neighboring ordinary satellites within a feasible inter-satellite link (ISL) range. The relay acts as the coordination node, responsible for partitioning a normalized partitionable  task and aggregating computed results. A fixed fraction $f$ 
of the task is constrained to be processed locally at the relay due to hardware, security, or mission-specific requirements, while the remaining fraction $(1-f)$ is split  among the relay and the $N$ neighboring satellites. Communication between the relay and each satellite occurs over dedicated ISLs with heterogeneous transmission rates, and all satellites exhibit heterogeneous processing speeds.

Under the MPCC assumption, the relay can simultaneously transmit load fractions to all neighboring satellites and concurrently receive their returned results once computation completes. The returned data from each satellite is modeled as a proportional fraction of the assigned task, capturing realistic scenarios such as feature aggregation, compressed sensing outputs, or partial inference results. The overall task completion time is defined as the maximum of the relay’s local computation time and the completion times of all satellite-assisted computations. Applying the DLT optimality criterion\cite{Bharadwaj1996Book}, the problem reduces to determining load fractions that equalize completion times across all participating nodes, leading directly to the closed-form expressions derived in the following sections.
\subsection{Task characteristics}
Modern DSS payloads increasingly execute on-board image processing, AI inference, signal and scientific data processing, SAR image analysis, and weather or environmental forecasting, exploiting space-edge computing to reduce downlink latency and bandwidth consumption. In such architectures, relay satellites primarily act as high-capacity intermediaries, supporting data relay, massive IoT aggregation, and telecom traffic optimization for space and ground users \cite{PhilabESA}. Many of these tasks are inherently parallelizable, motivating fine-grained task partitioning under application-specific constraints. For example, IoT aggregation across thousands of independent sensor streams can be decomposed into disjoint batches, while SAR and remote-sensing pipelines can partition imagery into spatial tiles with limited boundary coordination \cite{PhilabESA}. Work reported in \cite{SP2024}  presents a collaborative satellite computing framework that dynamically partitions a DNN workload into subtasks and distributes them across multiple satellites for parallel execution. A key contribution is the joint optimization of DNN partitioning and subtask-to-satellite assignment, ensuring efficient load balancing across the constellation. Also, the survey in \cite{QZ2024} systematically covers distributed satellite-based network architectures (including distributed regenerative payloads and multi-satellite collaborative transmission), key enabling technologies such as onboard AI, edge computing, and multi-source data fusion, as well as channel modeling and resource management frameworks. These characteristics make DSS a natural candidate for divisible-load-based scheduling and cooperative processing.

To capture this behavior, we introduce a parameter $\gamma$ to represent the data transfer needs of a task, defined as the fraction of the task that can be distributed from a root satellite to cooperating nodes during an off-loading decision. A high-$\gamma$ task, such as Monte Carlo simulations or pixel-level filtering, benefits substantially from parallel execution when inter-satellite links provide sufficient capacity, whereas low-$\gamma$ tasks with strong sequential dependencies incur communication overhead without meaningful latency reduction. By explicitly exposing $\gamma$, the task allocator can enforce application semantics while optimizing computation placement across heterogeneous satellites; for instance, a weather-processing task may retain global calibration centrally while offloading per-sensor checks to neighboring nodes \cite{Mark2023}. In practice, the effective usefulness of task distribution depends jointly on inter-satellite link capacity, task size, and computational intensity—factors that are now highly variable in modern LEO constellations employing both RF and optical ISLs.
\subsection{Notations and definitions}
We consider a single-level tree network, also referred to as a {\it star} network comprising a relay node as a root with $N$ children nodes as ordinary satellites. Let the relay node $0$ denote the root and nodes $i=1,\dots,N$ denote the children. We strictly follow the notations used in the DLT literature, defined here for continuity. The parameter $w_i$ denotes the \emph{computation time per unit load} at processor $i$ (so computing load fraction $\alpha$ at node $i$ takes time $\alpha w_i$). Similarly, the parameter $z_i$ denotes the \emph{communication time per unit load} on the link between the root and child $i$ (so transmitting load fraction $\alpha$ over that link takes time $\alpha z_i$). The task size $L$ is normalized to $1$. A fixed sequential fraction $f$ can be computed \emph{only} at the root. The remaining  part, defined earlier as $\gamma =(1-f)$ serves as the  divisible part will be split among the root and all the participating children:
\[
\alpha_0 + \sum_{i=1}^N \alpha_i = \gamma,\qquad \alpha_i \ge 0.
\]
The root computes both the mandatory portion $f$ and its share $\alpha_0$, i.e., the root computes total fraction $(f+\alpha_0)$.
\subsection{Parallel Communication and Deriving the Completion Times}
As mentioned in the introduction, we adopt a \emph{parallel (multi-port)} communication model in which the root can transmit load fractions to all children \emph{concurrently}. 
After computation, children transmit the results back to the root \emph{concurrently}, and the root can receive all such results concurrently. The result size returned by child $i$ is assumed to be a fraction $\beta$ of the assigned load size, i.e., result size is $\beta\cdot\alpha_i$, where $0<\beta<1$.
For child $i$, the forward transmission of load fraction $\alpha_i$ from root to child $i$ takes $\alpha_i z_i$; the computation at child $i$ takes $\alpha_i w_i$, and finally, the return transmission of results of size $k\alpha_i$ takes $\beta \cdot\alpha_i\cdot z_i$. Hence, the completion time of child $i$ (as observed at the root) is given by, 
\begin{equation}
T_i = \alpha_i z_i + \alpha_i w_i + \beta\alpha_i z_i
    = \alpha_i\bigl(w_i + (1+\beta)z_i\bigr).
\label{eq:Ti}
\end{equation}

\subsection{Optimality Criterion and Optimal Fractions}
The root computes fraction $(f+\alpha_0)$, so the root completion time is given by, 
\begin{equation}
T_0 = (f+\alpha_0)w_0.
\label{eq:T0}
\end{equation}
Thus, the overall finishing time (makespan) is
\begin{equation}
T = \max\{T_0, T_1, \dots, T_N\}.
\label{eq:makespan}
\end{equation}

A standard DLT optimality condition for divisible load problems (when all participating processors have positive load and no processor is idle at optimum) is the \emph{equal-finish-time} condition\cite{Robertazzi2003}:
\begin{equation}
T_0 = T_1 = \cdots = T_N = T^\star.
\label{eq:equalfinish}
\end{equation}

From (\ref{eq:Ti}) and (\ref{eq:equalfinish}), for each child $i=1,\dots,N$ with $\alpha_i>0$:
\begin{equation}
\alpha_i^\star = \dfrac{T^\star}{w_i + (1+\beta)z_i},\quad i=1,\dots,N.\;
\label{eq:alpha_i}
\end{equation}
From (\ref{eq:T0}) and (\ref{eq:equalfinish}):
\[
(f+\alpha_0^\star)w_0 = T^\star
\quad\Rightarrow\quad
\alpha_0^\star = \frac{T^\star}{w_0} - f.
\]
Thus,
\begin{equation}
\alpha_0^\star = \dfrac{T^\star}{w_0} - f
\label{eq:alpha_0}
\end{equation}

\subsection{Determining the Optimal Processing Time}
Using the constraint $\alpha_0 + \sum_{i=1}^N \alpha_i = \gamma$ together with (\ref{eq:alpha_i})--(\ref{eq:alpha_0}), we obtain:
\[
\left(\frac{T^\star}{w_0} - f\right) + \sum_{i=1}^N \frac{T^\star}{w_i + (1+k)z_i} = \gamma.
\]
With further algebraic steps, we obtain, 
\[
T^\star\left(\frac{1}{w_0} + \sum_{i=1}^N \frac{1}{w_i + (1+k)z_i}\right) = 1.
\]
Define:
\begin{equation}
S \;=\; \frac{1}{w_0} + \sum_{i=1}^N \frac{1}{w_i + (1+k)z_i}.
\label{eq:S}
\end{equation}
Then
\begin{equation}
T^\star = \frac{1}{S}
\label{eq:Tstar_case1}
\end{equation}

\subsection{Relay satellite's share: Feasibility condition}
The solution (\ref{eq:alpha_0}) requires $\alpha_0^\star \ge 0$. This imposes
\[
\frac{T^\star}{w_0} - f \ge 0
\quad\Longleftrightarrow\quad
f \le \frac{T^\star}{w_0}.
\]
Using (\ref{eq:Tstar_case1}):
\begin{equation}
f \le \frac{1}{w_0 S}
\label{eq:feasibility}
\end{equation}
This sets an upper bound on the extent to which the amount of sequential portion in the workload can be tolerated to allow the root to participate.  Now, we will derive the optimal solution. 
\subsection{Final Optimal Solution - Two cases}
Let us rewrite (\ref{eq:S}) as, 
\[
S=\frac{1}{w_0}+G,~ \text{where}~
G=\sum_{i=1}^N\frac{1}{w_i+(1+k)z_i}.
\]
\subsubsection*{Case 1: Root share is feasible ($\alpha_0^\star \ge 0$)}
If (\ref{eq:feasibility}) holds (equivalently $\alpha_0^\star\ge 0$), then the optimal makespan and fractions are: $T^\star=1/S$, 
$\alpha_i^\star= T^\star/(w_i+(1+k)z_i),\quad i=1,\dots,N\ \text{and}~ \alpha_0^\star=(T^\star/w_0)-f$
\subsubsection*{Case 2: Root-only fraction is too large (set $\alpha_0^\star=0$)}
If $\alpha_0^\star<0$ under the Case 1 formulas, then the root cannot take any part of $\gamma$ (beyond its mandatory $f$) and hence, we set $\alpha_0^\star = 0$. The remaining load $\gamma$ is split across children satisfying equal-finish among children. Thus, 
\begin{equation}
\alpha_i^\star=\gamma \left(\frac{\frac{1}{w_i+(1+k)z_i}}{\sum_{j=1}^N\frac{1}{w_j+(1+k)z_j}}\right),\quad i=1,\dots,N,~~ \alpha_0^\star=0
\label{eq:alpha_case2}
\end{equation}
In this regime, the overall makespan must accommodate both the root's mandatory computation $fw_0$ and the children completion time. Thus, the children equal-finish time for processing $\gamma$ is
\[
T_{\text{children}}=\frac{1-f}{G}.
\]
Therefore,
\begin{equation}
T^\star=\max\left(fw_0,\ \frac{\gamma}{G}\right)
\label{eq:Tstar_case2}
\end{equation}

{\it Remarks:} Under parallel (multi-port) forward and backward communications, there is no sequential transmission ordering coupling across children; consequently, the DLT optimality criterion yields simple closed-form expressions as above (subject to the feasibility check for $\alpha_0^\star$).
\section{Deadline Feasibility and Resource Sizing}
\label{sec:deadline}
In practical DSS operations, task execution is often subject to strict \emph{deadline constraints} arising from orbital visibility windows, sensing schedules, or real-time service requirements. Within the proposed MPCC-DLT framework, such requirements can be incorporated naturally by interpreting the target completion time as a deadline constraint, denoted by $T_{\mathrm{req}}$. A task is said to be feasible under the MPCC-DLT model if the optimal makespan $T^\star$ satisfies $T^\star \leq T_{\mathrm{req}}$.

From the closed-form MPCC-DLT solution, the optimal makespan for the case where the relay satellite participates in processing the divisible load is given by (\ref{eq:Tstar_case1}). 
Imposing the deadline constraint $T^\star \leq T_{\mathrm{req}}$ yields the feasibility condition:
\begin{equation}
\frac{1}{w_0}
+
\sum_{i=1}^{N}
\frac{1}{w_i + (1+\beta) z_i}
\;\geq\;
\frac{1}{T_{\mathrm{req}}}.
\label{eq:deadline_feasible}
\end{equation}

\subsection{Minimum Number of Cooperating Satellites}
Equation~(\ref{eq:deadline_feasible}) enables a direct derivation of the minimum number of neighboring satellites required to meet a given deadline. Define the \emph{effective service contribution} of satellite $i$ as
\begin{equation}
g_i \triangleq \frac{1}{w_i + (1+\beta) z_i}.
\label{eq:gi}
\end{equation}
Then, using (\ref{eq:Tstar_case1}), the deadline feasibility condition can be rewritten as
\begin{equation}
\sum_{i=1}^{N} g_i \;\geq\; 
\Delta
\label{eq:delta}
\end{equation}
 where, $\Delta \triangleq \left(\frac{1}{T_{\mathrm{req}}} - \frac{1}{w_0}\right)$. 
 It may be noted that $(1/T_{req})$ is the minimum total effective service rate needed to finish by the deadline and $1/w_0$ is the relay’s own effective service rate (if it were the only processor in the MPCC–DLT Case-1 in Section 2.7). Therefore, $\Delta$ is the rate deficit that must be supplied by the cooperating satellites. Thus, if $\Delta \leq 0$ then
 this leads to conclude $w_0 < T_{req}$, meaning the relay alone can meet the deadline (cooperation is not required). However, if $\Delta \geq 0$ the children must collectively contribute at least $\Delta$ to meet the deadline. The minimum number of cooperating satellites is therefore given by,
\begin{equation}
N_{\min}(T_{\mathrm{req}})
=
\min \left\{
N : \sum_{i=1}^{N} g_{(i)} \geq \Delta
\right\},
\label{eq:Nmin}
\end{equation}
where $g_{(1)} \ge g_{(2)} \ge \cdots$ denotes the ordered set of satellite contributions in descending order. This expression formalizes the intuitive requirement that satellites with higher processing speeds and higher-capacity inter-satellite links contribute more effectively toward meeting stringent deadlines.
\subsection{Aggregate Compute and Inter-Satellite Bandwidth Requirements}
The MPCC-DLT model reveals that the collective computation and communication resources enter the feasibility condition in an additive manner. Specifically, the left-hand side of (\ref{eq:deadline_feasible}) represents the \emph{aggregate effective service rate} of the relay-centered cluster. Satellites with large $w_i$ (slow processors) or large $z_i$ (low-rate inter-satellite links) contribute negligibly, whereas well-provisioned satellites provide substantial gains.

Rewriting $z_i = 1/R_i$, where $R_i$ denotes the inter-satellite link rate, the effective contribution becomes
\begin{equation}
g_i = \frac{1}{w_i + \frac{1+\beta}{R_i}}.
\label{eq:gi_rate}
\end{equation}
Equation~(\ref{eq:gi_rate}) explicitly captures the trade-off between on-board computation and communication bandwidth. In the computation-limited regime ($w_i \gg (1+\beta)/R_i$), increasing bandwidth yields diminishing returns, while in the communication-limited regime ($w_i \ll (1+\beta)/R_i$), improving inter-satellite link rates is essential to realize the benefits of distributed processing.
%
\section{Performance Evaluation}
\label{sec:performance}
In this section, we evaluate the proposed MPCC-DLT framework through simulation-based experiments that quantify its behavior under varying task sizes, application characteristics, and deadline-driven resource constraints. The evaluation focuses on relay-centric single-level DSS topologies and is designed to directly validate the analytical model developed earlier. We will first present results related to load allocation for static task instances, and then we will present real-time admission control results. 
\subsection{Experimental Parameters and Practical Interpretation}
Each simulation instance consists of a relay satellite and $N$ neighboring satellites connected via inter-satellite links (ISLs). For each satellite $i$, the parameter $w_i$ denotes the \emph{computation time per unit load} and is computed as
$w_i = (CI_{\rm task})/(CS_i)$ 
where, $CI_{\rm task}$ is the compute intensity of the task, in Flops/MB and $CS_i$ is the on-board processing throughput (compute speed) of the node $i$, in Flops/sec. Thus, 
    assigning $L_i$ amount of task results in an actual computation time of $L_i\cdot w_i$ seconds. This formulation captures realistic heterogeneity in satellite payload processors, which can range from low-power embedded processors to more capable AI accelerators especially in edge-based satellite systems\cite{Mark2023}.

Similarly, the communication parameter $z_i$ represents the {\it communication time per unit load} on the ISL between the relay and satellite $i$, defined as $z_i = 1/ (\text{ISL~bandwidth})$
Thus, transmitting $L_i$ amount of task requires $L_i \cdot z_i$ seconds. The return of computed results is explicitly modeled using a result-size ratio $\beta$, such that result transmission requires $\beta L_i \cdot z_i$ seconds. This abstraction accommodates both RF and optical ISLs, whose data rates in modern LEO constellations can span several orders of magnitude \cite{AUChaud2023}.
\vspace{-0.4cm}
\subsection{Application Task Types and Typical Parameter Ranges}
Rather than repeating qualitative descriptions, Table~\ref{tab:task_params} summarizes the representative application classes considered in the evaluation along with the typical parameter ranges used in the simulations. These ranges are motivated by recent DSS and space-edge computing studies \cite{AYBoucetta2024,SP2024, PhilabESA} and are chosen to reflect realistic operational conditions.
\begin{table}[t]
\centering
\caption{Parameter ranges for major satellite tasks}
\label{tab:task_params}
\setlength{\tabcolsep}{3pt}
\begin{tabular}{lccccc}
\toprule
\textbf{Parameter} &
\textbf{IoT Agg.} &
\textbf{AI Inf.} &
\textbf{Img./Sig. Pre.} &
\textbf{Sci. Data} \\
\midrule
L(MB) &
$10^2$--$10^3$ &
$10^2$--$10^4$ &
$10^3$--$10^4$ &
$10^3$--$10^5$ \\

CI(Flops/MB) &
$10^6$--$10^7$ &
$10^8$--$10^9$ &
$10^7$--$10^8$ &
$10^8$--$10^{10}$ \\

\boldmath$\gamma$ &
$0.6$--$0.8$ &
$0.7$--$0.9$ &
$0.5$--$0.7$ &
$0.4$--$0.6$ \\

\boldmath$\beta$ &
$0.05$--$0.15$ &
$0.1$--$0.3$ &
$0.05$--$0.2$ &
$0.1$--$0.25$ \\
\bottomrule
\end{tabular}
\end{table}
The parameter $\gamma$ specifies the fraction of task that can be offloaded from the relay to neighboring satellites, while $\beta$ captures the relative size of the returned results. High-$\gamma$ tasks correspond to embarrassingly parallel task, whereas lower-$\gamma$ tasks retain a larger sequential component(non-divisible) at the relay. Let us now present our results from different class of experiments. 

\subsection{Effect of Task Scaling with Fixed Application Semantics}
\begin{figure}
     \centering
     \includegraphics[width=1\linewidth]{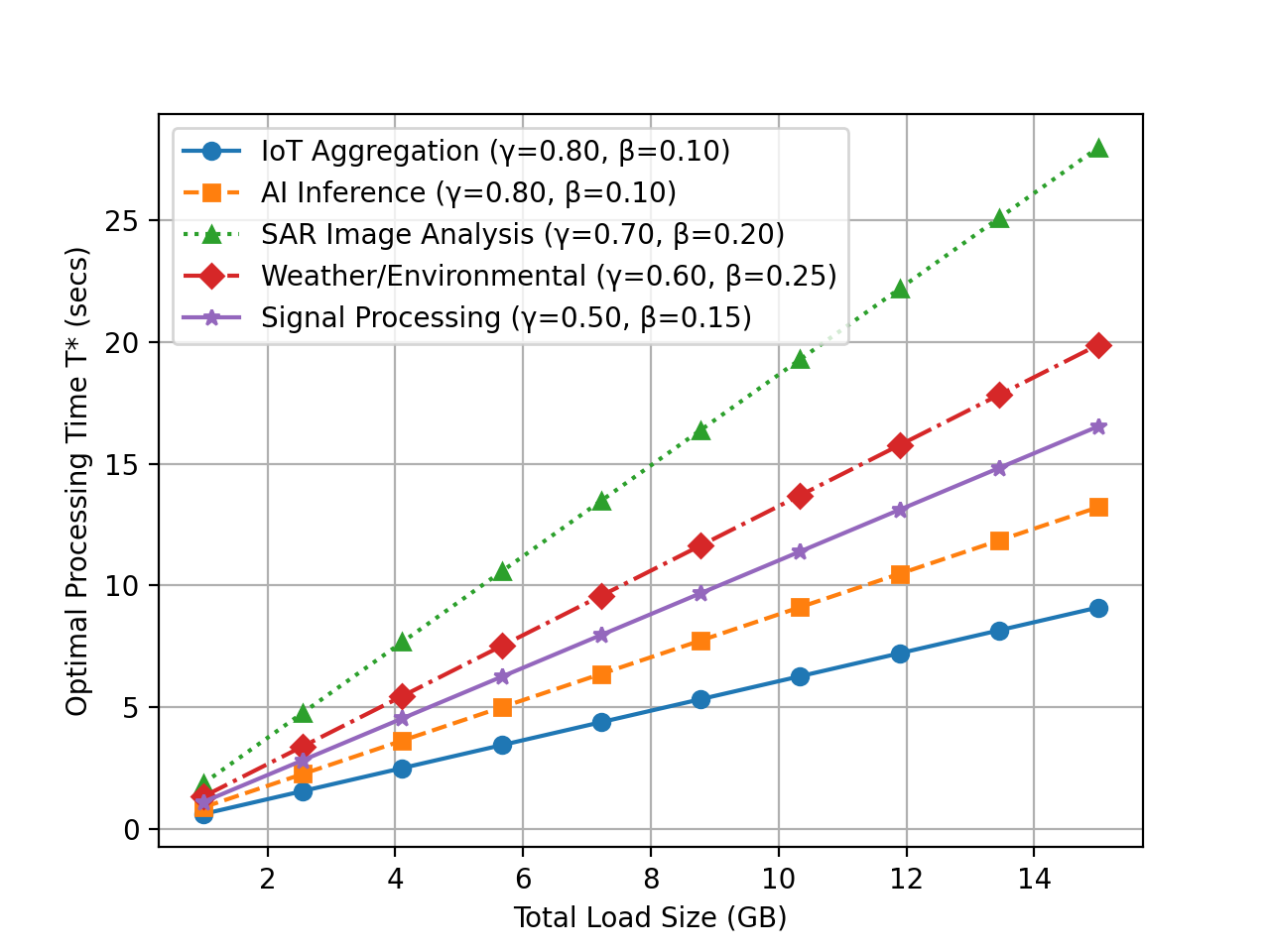}
     \caption{Effect of task Scaling with Fixed Application Semantics}
     \label{fig:expt1}
 \end{figure}

In this experiment, we examine the effect of increasing task size while retaining application semantics fixed across multiple workload classes. For each application, the parameter $\gamma$ and the result ratio $\beta$ are set to representative constant values, as summarized in Table~\ref{tab:task_params}.  Also, the total data size is scaled across several runs. The effect of task size on the ideal completion time $T^\star$ under the same platform conditions is isolated in this design. As illustrated in Fig.~\ref{fig:expt1}, $T^\star$ increases linearly with data size for all application types, including AI inference, IoT aggregation, SAR image analysis, weather processing, and signal processing. The differing slopes across applications capture variations in compute intensity and communication overhead, while the linear trend confirms the scaling behavior predicted by the MPCC--DLT model. Fixing application semantics ensures that the observed trends arise solely from load scaling rather than changes in task structure or parallelizability.

\subsection{Evaluating the Sensitivity to Application Characteristics}
In this experimental study, we evaluate the sensitivity of MPCC-DLT performance to application-level variability. Thus, the task size is maintained within a comparable range, while task type, compute intensity, result-size ratio $\beta$, and the parameter $\gamma$ are varied across runs according to the ranges in Table~\ref{tab:task_params}. In contrast to the earlier experiment, $\gamma$ is randomly sampled within task-specific ranges to capture realistic variability in partitionability across inputs and operating conditions. Thus, we see that this experiment demonstrates that tasks with higher sampled $\gamma$ and lower $\beta$ benefit most from multi-port concurrent execution, achieving lower completion times, while communication-heavy or weakly parallel tasks experience reduced gains. The results, summarized in Table~\ref{tab:expt2}, show that the MPCC-DLT framework adapts gracefully to heterogeneous task without requiring task-specific tuning.
\begin{table}
\centering
\caption{Parameter variation summary}
\label{tab:expt2_params}
\begin{tabular}{|l|l|}
\hline
\textbf{Parameter} & \textbf{Variation Strategy} \\ \hline
Task type & Randomly selected per run \\ \hline
Task size & Sampled within task range \\ \hline
Compute intensity & Sampled within task range \\ \hline
Distributability $\gamma$ & Randomly sampled in task range \\ \hline
Result ratio $\beta$ & Randomly sampled in task range \\ \hline
Satellite compute speeds & Fixed per topology (heterogeneous) \\ \hline
ISL bandwidths & Fixed per topology (heterogeneous) \\ \hline
\end{tabular}
\end{table}
\begin{table}[t]
\centering
\small
\setlength{\tabcolsep}{1pt}
\renewcommand{\arraystretch}{0.97}
\caption{$T^{*}$ for different application tasks (averaged over 12 runs for each application task with varying CI, $\gamma$, and $\beta$)}
\label{tab:expt2}
\begin{tabular}{lccccc}
\toprule
\textbf{Application} &
\textbf{TaskSize (GB)} &
\boldmath$\gamma$ &
\boldmath$\beta$ &
\textbf{CI.$10^7$} &
$T^{*}$ \textbf{(s)} \\
\midrule
IoT Aggrg. &
4.5084 &
0.7853 &
0.1415 &
4.5319 &
2.7084 \\
AI Inference &
5.0053 &
0.7098 &
0.1705 &
12.791 &
6.1624 \\
SAR-Image Proc. &
8.3039 &
0.6981 &
0.12209 &
45.748 &
27.324 \\
Weather/Envt. &
5.4324 &
0.53470 &
0.18132 &
38.144 &
23.688 \\
Signal Proc. &
6.6729 &
0.5155 &
0.1411 &
30.098 &
22.2945 \\
\bottomrule
\end{tabular}
\end{table}

In contrast to an earlier experiment, here, the parameter $\gamma$ is not fixed but sampled within task-dependent ranges to capture application variability.  Several important trends emerge from the results of this experiment. First, tasks with higher sampled values of $\gamma$ consistently achieve lower completion times $T^\star$, confirming that increased parallelizability directly improves the effectiveness of multi-port concurrent execution. This trend is particularly pronounced for AI inference and simulation-style tasks, where a large fraction of the computation can be offloaded without introducing strong dependencies. 

\begin{figure}
    \centering
    \includegraphics[width=1\linewidth]{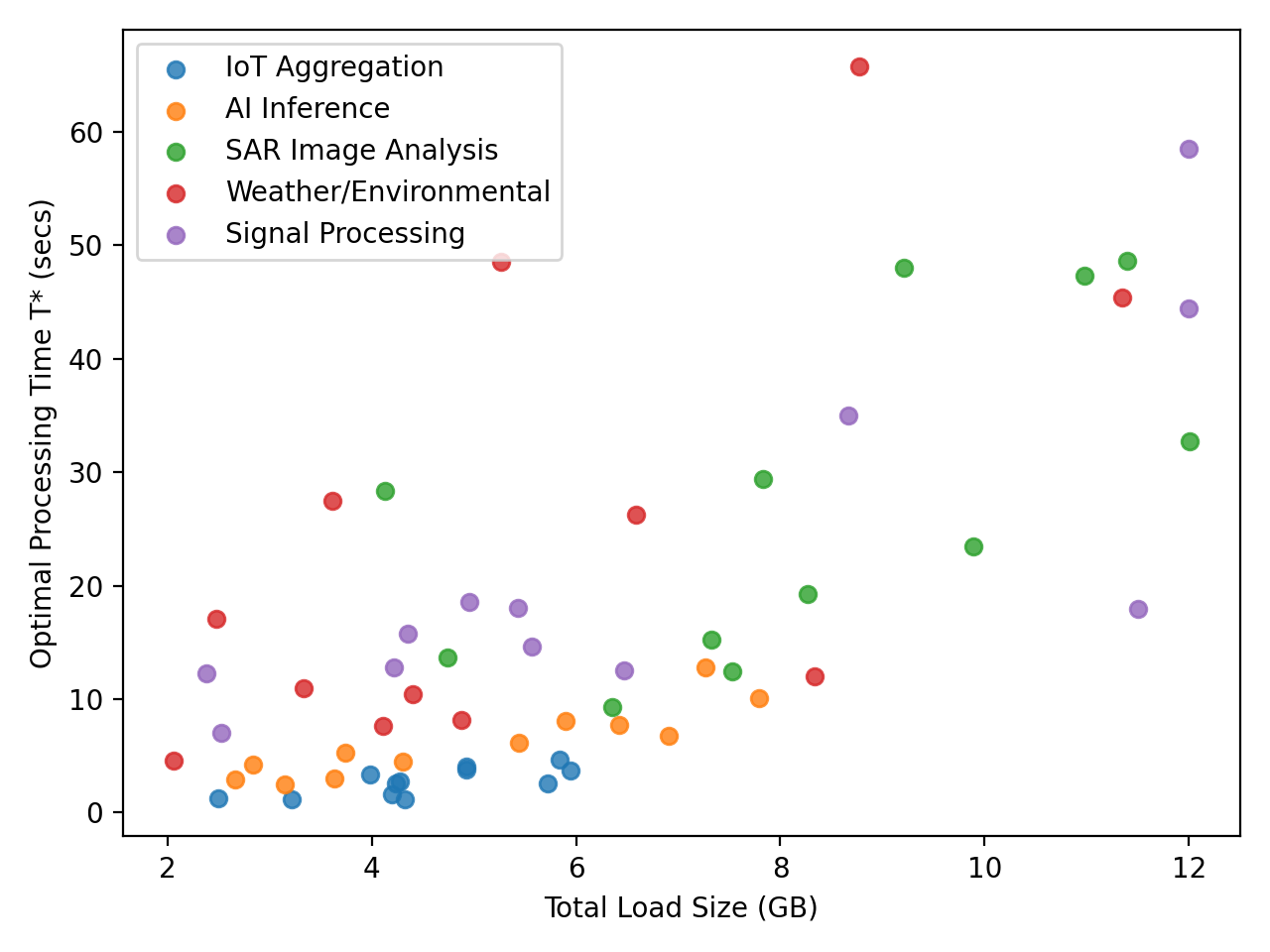}
    \caption{Task size versus $T^\star$}
    \label{fig:scatter}
\end{figure}

The scatter plot shown in Fig. \ref{fig:scatter} shows the relationship between the task size and the $T^\star$ under the MPCC–DLT model for different application types. Each point corresponds to one simulation run, with colors indicating the task category. Since application semantics are different and fixed, this plot highlights the variability introduced by heterogeneous task characteristics such as compute intensity, $\gamma$ and result-size ratio $\beta$. The dispersion of points for a given task size shows that tasks with higher parallelizability (e.g., IoT aggregation and AI inference) consistently achieve lower completion times, while compute and communication-intensive tasks (e.g., SAR image analysis and weather/environmental processing) incur significantly higher $T^\star$ even at comparable task sizes. The trend visually demonstrates that task size alone is insufficient to predict performance in distributed satellite systems; instead, application semantics play a dominant role, thus validating the need for application-aware task allocation, as enabled by the MPCC–DLT framework, and provides intuition for why a single scheduling policy cannot optimally serve all DSS tasks.
Second, the result-size ratio $\beta$ plays a critical moderating role. Even for tasks with high $\gamma$, larger values of $\beta$ increase the communication overhead associated with result return, thereby reducing the net benefit of distributed processing. This effect is especially visible in bandwidth-constrained scenarios, where the return phase becomes a dominant component of the overall completion time. These observations validate the analytical role of $(1+\beta)z_i$ in the MPCC-DLT formulation.

Third, the results demonstrate that MPCC-DLT adapts gracefully to heterogeneous task characteristics without requiring task-specific tuning. Completion times vary smoothly across runs as $\gamma$, compute intensity, and $\beta$ change, indicating that the task allocation automatically balances computation and communication according to instantaneous task semantics. This robustness is a key advantage for DSS deployments, where task characteristics could vary significantly across sensing modes, mission phases, or environmental conditions.

On the whole, our results highlight the importance of jointly considering task distributability, computational intensity, and result-return overhead when designing cooperative satellite processing strategies. The observed trends confirm that MPCC-DLT provides not only optimal task allocation under fixed assumptions, but also predictable and interpretable behavior under realistic application variability, making it well suited for practical DSS operation and planning.
\subsection{On Deadline-Driven Resource Sizing}
Now we will attempt to validate the analytical deadline feasibility and resource sizing conditions derived earlier in Section 2. The objective of this experiment is to determine the minimum level of cooperative resources required to meet a prescribed task completion deadline. Specifically, the target deadline $T_{\mathrm{req}}$ is fixed, and the number of cooperating satellites is progressively increased until the analytical feasibility condition is satisfied. In our simulation runs, $T_{\rm req}$ is set to $0.6\,w_0$, where $w_0$ denotes the relay’s computation time per unit task, so that the relay alone cannot meet the deadline.

\begin{figure}
    \centering
    \includegraphics[width=1\linewidth]{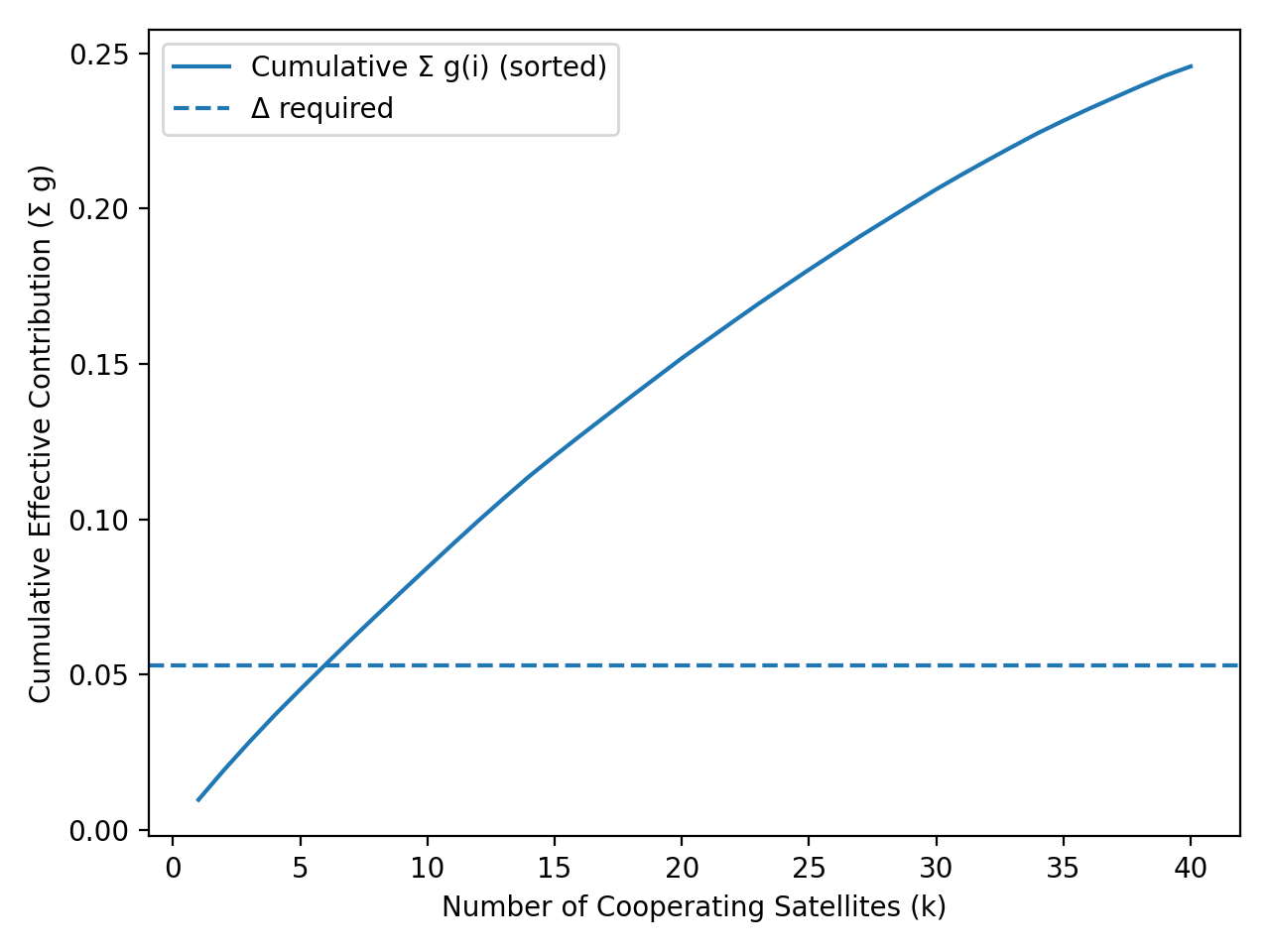}
    \caption{Relationship between the number of cooperating satellites and the cummulative effective contribution}
    \label{fig:expt3}
\end{figure}
Figure~\ref{fig:expt3} illustrates the relationship between the number of cooperating satellites and the resulting optimal completion time. The y-axis represents the optimal completion time $T^\star$ obtained from the MPCC-DLT allocation, which corresponds to the maximum finishing time across the relay and all cooperating satellites. This metric directly captures whether the distributed execution can meet the imposed deadline constraint.

The horizontal dotted line in Fig.~\ref{fig:expt3} denotes the target deadline $T_{\rm req}$. This line serves as a feasibility threshold: The y-axis represents the cumulative effective service contribution of the relay-centered cluster, defined as the reciprocal of the optimal completion time under the MPCC–DLT allocation. The horizontal dotted line corresponds to the reciprocal of the target deadline, $1/T_{req}$. Configurations whose cumulative effective contribution exceeds this threshold satisfy the deadline constraint, whereas configurations below the line violate it. The intersection point between the $T^\star$ curve and the deadline line therefore identifies the minimum number of cooperating satellites required to satisfy the deadline.

Several important trends can be observed. First, $T^\star$ decreases monotonically as the number of cooperating satellites increases, reflecting the additive contribution of additional computational and communication resources under the MPCC-DLT model. Second, the marginal reduction in completion time diminishes as more satellites are added, indicating diminishing returns once the aggregate service capacity becomes sufficiently large. This behavior is consistent with the analytical effective service contribution derived earlier and highlights that not all additional satellites contribute equally, particularly when their processing speeds or inter-satellite link capacities are limited.

The significance of this experiment lies in its direct applicability to DSS planning and admission control. By identifying the smallest satellite subset that satisfies a given deadline, MPCC-DLT provides a principled mechanism for determining when cooperative processing is necessary and when local execution is sufficient. Moreover, the results demonstrate that meeting stringent deadlines can be achieved either by increasing the number of participating satellites or by improving on-board computation and ISL bandwidth, thereby enabling flexible tradeoffs between constellation size, hardware capability, and mission requirements.
\vspace{-0.3cm}
\section{Real-Time Admission Control for MPCC-DLT in DSS}
\subsection{Motivation and Objectives}
While the MPCC-DLT framework provides closed-form optimal task  allocation for static task instances, practical DSSs operate under \emph{real-time task arrivals} with strict deadline constraints. In such settings, not all arriving tasks can be admitted simultaneously due to finite onboard compute and inter-satellite communication resources. To bridge this gap between theory and practice, we extend the MPCC-DLT framework with an \emph{admission control mechanism} that explicitly determines whether an arriving task can be scheduled without violating its deadline. The primary objective of this real-time extension is to evaluate how MPCC-DLT behaves under stochastic arrivals and to quantify its ability to meet latency guarantees in dynamic DSS environments.
\subsection{Real-Time Simulation Framework}
We consider a SLTN abstraction, where a relay satellite cooperates with a set of neighboring satellites within inter-satellite link (ISL) range. The platform consists of $N=13$ satellites (one relay and twelve neighbors), representing a moderately sized cooperative cluster typical of contemporary low Earth orbit (LEO) DSS formations. Each satellite is assigned a heterogeneous processing speed $w_i \in [0.02, 0.08]$, while ISL communication speeds are sampled from $z_i \in [0.01, 0.06]$, reflecting realistic variability in onboard processors and RF/optical crosslinks.

Task arrivals follow a Poisson process, capturing asynchronous sensing, inference, and data-processing tasks generated onboard or relayed from other satellites. Upon arrival, a task is either admitted or blocked based on a feasibility test derived from MPCC-DLT: A task is accepted only if its predicted completion time, accounting for current resource occupancy, does not exceed its deadline. This admission control policy allows us to directly study \emph{blocking probability} as a performance metric, which is critical for real-time DSS schedulers.
\subsection{Task Classes and Parameterization}
To ensure relevant demonstrations, we selected four representative DSS task classes spanning a range of distributability and communication characteristics. Each task is characterized by a distributability parameter $\gamma$, a result-to-input ratio $\beta$, and a random task size multiplier $L$ to capture task variability. Table~\ref{tab:rt_tasks} summarizes the task parameters used in the experiments.
\begin{table}[t]
\centering
\caption{DSS task Classes for Real-Time MPCC-DLT Experiments}
\label{tab:rt_tasks}
\begin{tabular}{lccc}
\hline
\text{Task Class} & $\boldsymbol{\gamma}$ & $\boldsymbol{\beta}$ & \text{Example Applications} \\
\hline
Class A  & 0.8 & 0.10 & Telemetry fusion, sensor aggregation \\
Class B   & 0.8 & 0.20 & CNN inference, object detection \\
Class C & 0.6 & 0.40 & SAR image tiling and filtering \\
Class D    & 0.35 & 0.10 & Control decoding, transactional tasks \\
\hline
\end{tabular}
\end{table}
where classes A-D are application tasks that belong to IoT Aggregation, AI Inference, SAR Processing, and Low-$\gamma$, respectively. The chosen parameter ranges align very well with the reported DSS tasks and ensure that both computation-dominated and communication-dominated regimes are exercised.
\subsection{Deadline and Offered-Load Design}
A key design choice in the real-time experiments is the definition of task deadlines and arrival intensity. For each task instance, the deadline is set as $T_{\rm req} = (1+\delta)\,S$
where $S$ is the MPCC-DLT predicted service time for that task in isolation and $\delta>0$ is a slack parameter. This ensures that tasks are feasible when admitted alone, while congestion-induced delays can still trigger deadline violations. 

Our arrival rates are parameterized through an \emph{offered- load} $a = \lambda\,\mathbb{E}[S]$, where $\lambda$ is the Poisson arrival rate. Using normalized offered-load levels $a \in \{0.3,\,0.7,\,1.2\}$ allows us to systematically explore {\it lightly loaded, moderately loaded, and overloaded regimes} independent of absolute service times. This normalization is particularly useful for DSS, where task characteristics can vary significantly across missions.
\vspace{-0.2cm}
\subsection{Experimental Cases and Intended Insights}
We design a minimal yet robust experimental cases to isolate the causal impact of key MPCC-DLT parameters. Specifically, we evaluate blocking probability as a function of: (i) offered-load $a$ across all task classes, (ii) non-divisible fraction $\gamma$ to expose Case-1 and Case-2 operating regimes captured in Section 2, and (iii) ISL bandwidth scaling to assess communication sensitivity. This structured design enables clear attribution of observed trends to distributability parameter $\gamma$, result ratio $\beta$, and local processing constraints, thereby demonstrating how MPCC-DLT can inform real-time scheduler design for DSS.
\subsection{Results Interpretation and Design Insights}
\label{RIDI}
We will now analyze our real-time MPCC-DLT results obtained under the admission control framework described above. Our  focus is on understanding how blocking probability evolves with offered-load, task structure, and inter-satellite communication capacity, and on extracting actionable insights for DSS scheduler design.
\vspace{-0.3cm}
\subsubsection*{\ref{RIDI}.1 Blocking Behavior Under Increasing Offered-Load}
\begin{figure}
    \centering
    \includegraphics[width=1\linewidth]{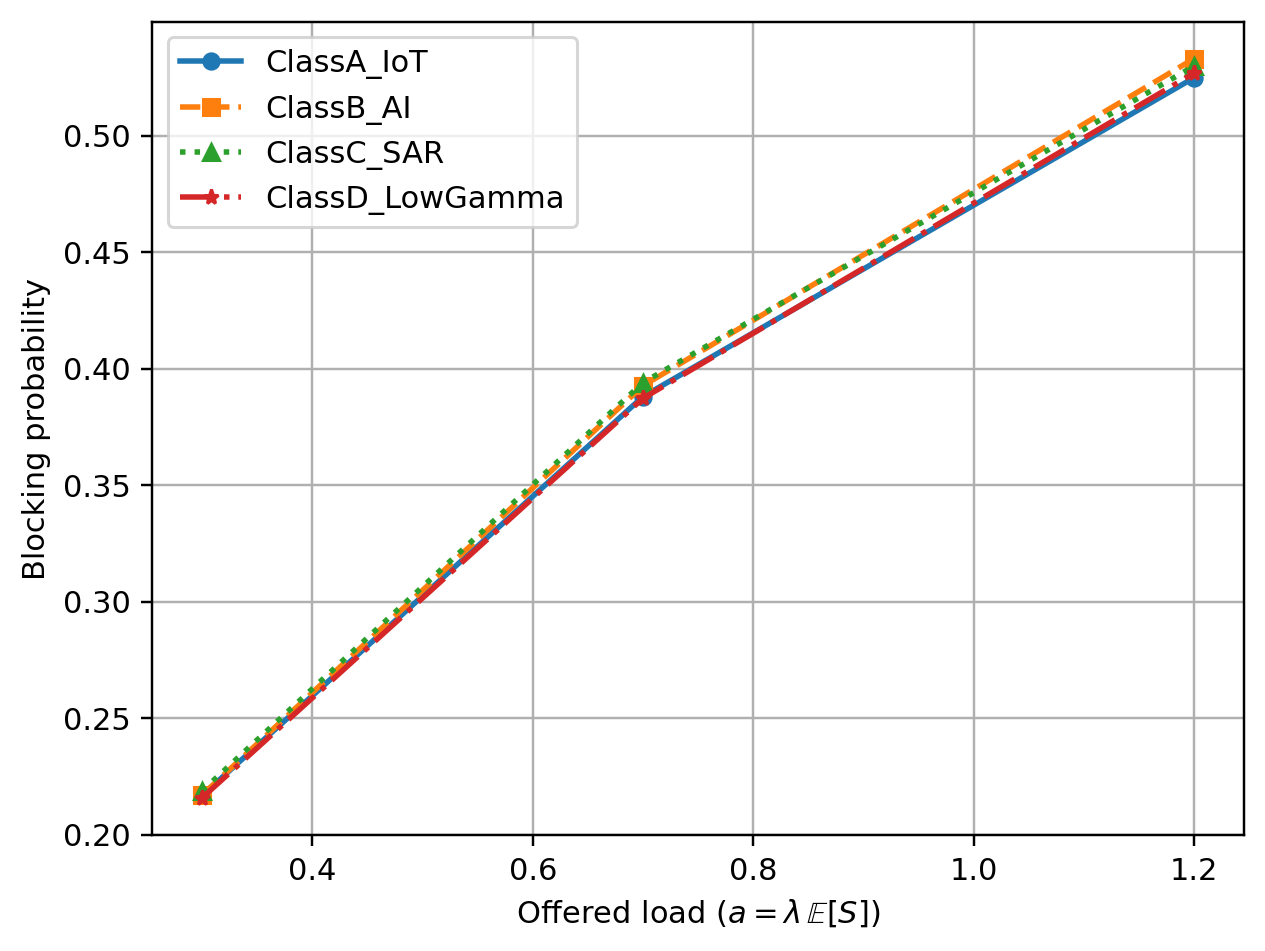}
    \caption{Blocking versus OfferedLoad}
    \label{fig:Blocking_vs_OfferedLoad_RTv2}
\end{figure}

Fig.~\ref{fig:Blocking_vs_OfferedLoad_RTv2} shows the blocking probability as a function of normalized offered-load $a=\lambda\mathbb{E}[S]$ for the four representative task classes. As expected, blocking probability increases monotonically with offered-load across all task types. However, the rate of increase differs significantly across tasks. Highly distributable tasks (Classes~A and~B with $\gamma=0.8$) exhibit substantially lower blocking at moderate load ($a=0.7$) compared to communication-intensive or low-distributability tasks (Classes~C and~D). The apparent cluttering of curves arises from the use of normalized offered-load, which aligns tasks at comparable utilization levels, and from task-size variability introduced to reflect realistic DSS operation. 
This result highlights the fact that MPCC-DLT not only reduces nominal service time but also improves \emph{admission resilience} under congestion. From a system design perspective, this implies that prioritizing highly distributable tasks during peak load can significantly improve deadline satisfaction without increasing hardware resources.
\subsubsection*{\ref{RIDI}.2 Impact of Sequential Fraction and Case-1/Case-2 Transition}
\begin{figure}
    \centering
    \includegraphics[width=1\linewidth]{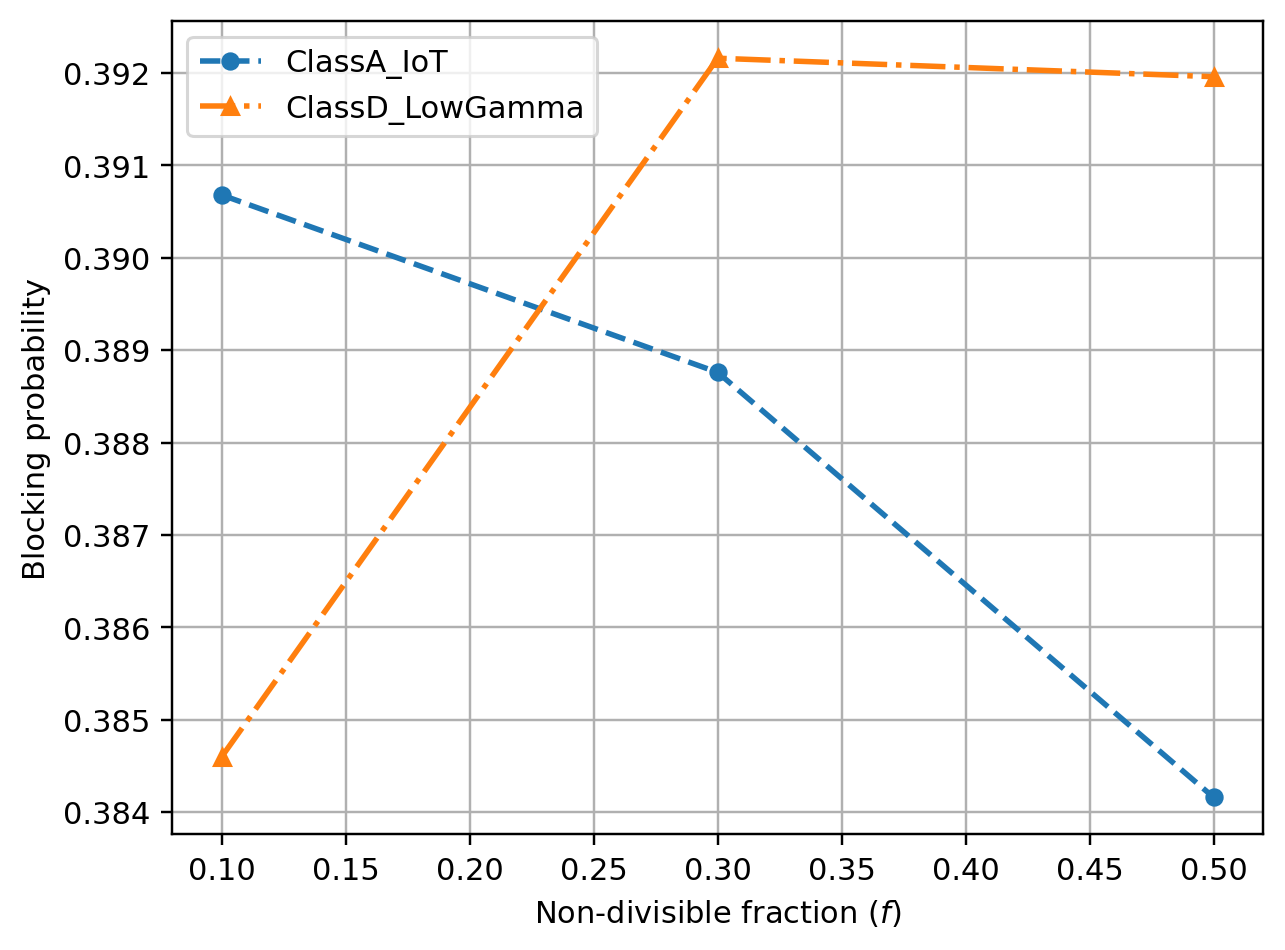}
    \caption{Effect of Sequential Part on Blocking for Class A and D tasks}
    \label{fig:Blocking_vs_f_RTv2}
\end{figure}

Fig.~\ref{fig:Blocking_vs_f_RTv2} illustrates blocking probability as a function of the sequential (non-divisible) fraction $f$ for a high-$\gamma$ tasks (Class~A) and a low-$\gamma$ tasks (Class~D) at fixed offered-load $a=0.7$. For both tasks, blocking probability increases with $f$, but the degradation is much steeper for the low-$\gamma$ class.
This trend reflects the transition from Case-1 (fully MPCC-dominated execution) to Case-2 behavior, where root-only computation becomes the bottleneck. The results confirm that even when sufficient resources are available, large non-divisible components can negate the benefits of cooperative processing. For designers, this underscores the importance of exposing task structure parameters (such as $f$) to the scheduler, rather than relying solely on aggregate task size.
\subsubsection*{\ref{RIDI}.3 Sensitivity to Inter-Satellite Bandwidth}
\begin{figure}
    \centering
    \includegraphics[width=1\linewidth]{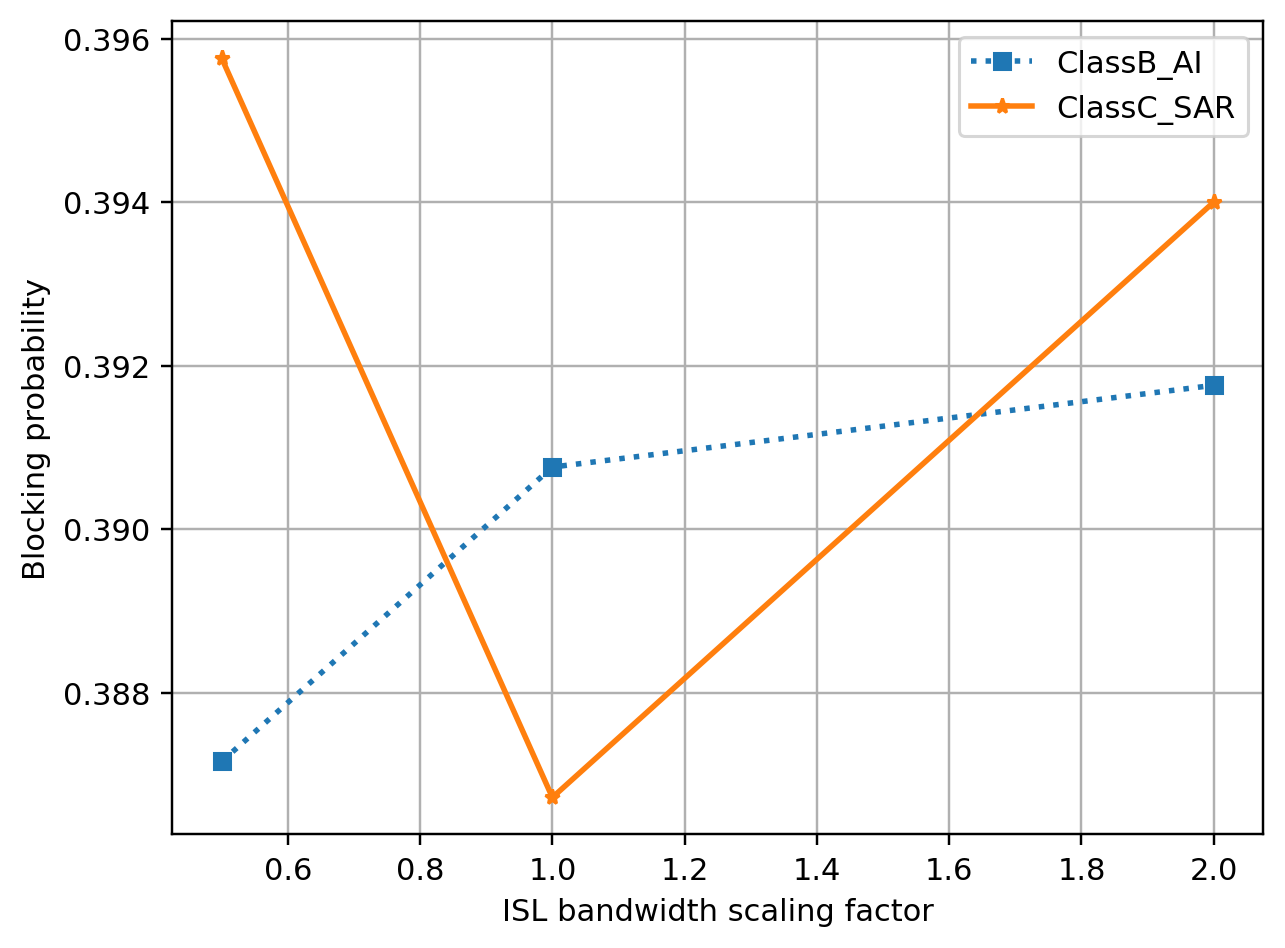}
    \caption{Effect of ISL Bandwidth scaling on Blocking Probability}
    \label{fig:Blocking_vs_BWscale_RTv2}
\end{figure}

In Fig.~\ref{fig:Blocking_vs_BWscale_RTv2} we examine the effect of ISL bandwidth scaling on blocking probability for Class~B (AI inference) and Class~C (SAR processing). Increasing ISL bandwidth significantly reduces blocking for the communication-heavy SAR task, while the improvement is more modest for AI inference tasks.
This differential sensitivity indicates that MPCC-DLT can guide \emph{bandwidth-aware task placement} - This means, tasks with high result ratios $\beta$ benefit disproportionately from higher ISL capacity. Consequently, future DSS architectures can exploit this insight by co-designing ISL upgrades with expected task mixes, rather than over-provisioning bandwidth uniformly.
%
\section{Conclusions}
In this paper, we proposed the first MPCC-enabled DLT formulation for relay-centric DSS, capturing concurrent data dissemination, parallel computation, and result return under heterogeneous on-board processing capabilities and inter-satellite link conditions. The proposed framework is shown to yield closed-form expressions that explicitly quantify how computation speed, link bandwidth, and result-size ratio jointly determine the optimal completion time through an aggregate effective service contribution. This analytical tractability provides a rigorous foundation for evaluating cooperative satellite processing under realistic multi-port communication model.

Our results reveal that highly distributable tasks ($\gamma \geq 0.7$) can achieve substantial latency reduction through cooperation, while communication-heavy tasks exhibit diminishing returns as result-size overhead increases. More importantly, the derived deadline feasibility conditions expose a direct and interpretable relationship between task urgency, collective compute capability, and inter-satellite bandwidth, enabling explicit sizing of cooperative satellite clusters. These insights move beyond performance characterization and establish MPCC-DLT as a practical design and decision-making tool for time-critical distributed satellite operations, addressing a gap not previously resolved in the DSS and satellite edge computing literature.

From a task scheduler design perspective, the results highlight the importance of application-aware scheduling that jointly considers task distributability, compute intensity, and result return cost. The MPCC-DLT framework enables schedulers to make informed decisions on when to offload, how many satellites to engage, and when local execution is sufficient. From a system standpoint, the proposed model provides actionable guidelines for relay satellite selection, ISL capacity provisioning, enabling DSS operators to balance latency, bandwidth utilization, and on-board equivalent energy consumption in a systematic manner. As such, the framework offers a unifying analytical tool for both algorithmic scheduler design and system-level DSS planning. 

Our MPCC-DLT admission control framework demonstrates that blocking probability in DSS is governed not only by arrival intensity but also by intrinsic task properties captured by $(\gamma,\beta,f)$. Our framework provides a principled mechanism to predict and control deadline violations under real-time operation. For system designers, the key takeaway is that meaningful latency guarantees can be achieved through \emph{task-aware scheduling} and \emph{selective cooperation}, even on moderately sized satellite clusters, without resorting to excessive compute or communication over-provisioning. An immediate future extension to this model is to include the influence of other resources and dynamic formation of satellite constellations such as mesh networks as in practice. In addition, motivated by the work reported in \cite{AUChaud2023}, we can attempt to study the trade-off between longer link distances and routing efficiency as this can provide design insights for latency-optimized LEO constellations. 
\section*{ACKNOWLEDGMENTS}
The author acknowledges the use of an integrated AI {\bf Writefull} tool embedded in the Overleaf Pro (LaTex) version to improve clarity in writing for language editing and figures/tables formatting assistance. Further, AI tool ChatGPT v5.2 was used to identify certain relevant  references([4],[17],[18]). The contents of these references have been independently verified and appropriately cited. The author retains full responsibility for the content and interpretations presented in this work.
\begingroup
\sloppy
\setlength{\parindent}{0pt}
\setlength{\parskip}{2pt}

\endgroup
\end{document}